%%%%%%%%%%%%%%%%%%%%%%%%%%%%%%%%%%%%%%%%%%%%%%%%%%%%%%%%%%%%%%%%%%%%%%%%%%%%
%%
%%   ChiLin.tex  by   Tristan Hubsch   thubsch@howard.edu
%%  (TeX+harvmac)
%%
%%%%%%%%%%%%%%%%%%%%%%%%%%%%%%%%%%%%%%%%%%%%%%%%%%%%%%%%%%%%%%%%%%%%%%%%%%%%
 %
 % Formatting macros
\input harvmac
%%%%%%%%%%%%%%%%%%%%%%%%%%%%%%%%%%%%%%%%%%%%%%%%%%%%%%%%%%%%%%%%%%%%%%%%
%
%  "zyp.tex", a set of macros to be used after loading "harvmac.tex"
%   Latest change: 25. X '98., by Tristan Hubsch
%
%%%%%%%%%%%%%%%%%%%%%%%%%%%%%%%%%%%%%%%%%%%%%%%%%%%%%%%%%%%%%%%%%%%%%%%%
 %
\catcode`@=11
\def\rlx{\relax\leavevmode}                   % Guess what this does...
 %
 %
%%%%%%%%%%%%%%%%%%%%%%%%%%%%%%%%%%%%%%%%%%%%%%%%%%%%%%%%%%%%%%%%%%%%%%%%
%%%****FONTS****FONTS****FONTS****FONTS****FONTS****FONTS****FONTS****%%
 % 
 % One point less than "\ninepoint", defined by "harvmac.tex"
\font\eightrm=cmr8 \font\eighti=cmmi8 \font\eightsy=cmsy8
  
\skewchar\eighti='177 \skewchar\eightsy='60
%

 % 
 % Bold-face fonts in math
\font\tenmib=cmmib10
\font\sevenmib=cmmib10 at 7pt % =cmmib7 % if you have it
\font\fivemib=cmmib10 at 5pt  % =cmmib5 % if you have it
\font\tenbsy=cmbsy10
\font\sevenbsy=cmbsy10 at 7pt % =cmbsy7 % if you have it
\font\fivebsy=cmbsy10 at 5pt  % =cmbsy5 % if you have it
\def\BMfont{\textfont0\tenbf \scriptfont0\sevenbf
                              \scriptscriptfont0\fivebf
            \textfont1\tenmib \scriptfont1\sevenmib
                               \scriptscriptfont1\fivemib
            \textfont2\tenbsy \scriptfont2\sevenbsy
                               \scriptscriptfont2\fivebsy}
\def\BM#1{\rlx\ifmmode\mathchoice
                      {\hbox{$\BMfont#1$}}
                      {\hbox{$\BMfont#1$}}
                      {\hbox{$\scriptstyle\BMfont#1$}}
                      {\hbox{$\scriptscriptstyle\BMfont#1$}}
                 \else{$\BMfont#1$}\fi}
 %
 % If you don't have the above fonts, comment out the above 21 lines,
 % >> send two pounds of live cockroaches to your TeX distributor <<
 % and use the Poor man's boldface (a la D. Knuth, and a bit better):
 %
 %\def\BM#1{\relax\leavevmode\setbox0=\hbox{$#1$}
 %           \kern-.025em\copy0\kern-\wd0
 %            \kern.05em\copy0\kern-\wd0
 %             \kern-.025em\raise.0433em\copy0\kern-\wd0
 %              \raise.0144em\box0 }
 %
 % Some black-board bold letters (alternatives to those in "msbm10")
 % ...should work rather well in sub- and super-scripts also...
 %
\def\inbar{\vrule height1.5ex width.4pt depth0pt}
\def\sinbar{\vrule height1ex width.35pt depth0pt}
\def\ssinbar{\vrule height.7ex width.3pt depth0pt}
\font\cmss=cmss10
\font\cmsss=cmss10 at 7pt
\def\ZZ{\rlx\leavevmode
             \ifmmode\mathchoice
                    {\hbox{\cmss Z\kern-.4em Z}}
                    {\hbox{\cmss Z\kern-.4em Z}}
                    {\lower.9pt\hbox{\cmsss Z\kern-.36em Z}}
                    {\lower1.2pt\hbox{\cmsss Z\kern-.36em Z}}
               \else{\cmss Z\kern-.4em Z}\fi}
\def\Ik{\rlx{\rm I\kern-.18em k}}  % Yes, I know. This ain't capital.
\def\IC{\rlx\leavevmode
             \ifmmode\mathchoice
                    {\hbox{\kern.33em\inbar\kern-.3em{\rm C}}}
                    {\hbox{\kern.33em\inbar\kern-.3em{\rm C}}}
                    {\hbox{\kern.28em\sinbar\kern-.25em{\sevenrm C}}}
                    {\hbox{\kern.25em\ssinbar\kern-.22em{\fiverm C}}}
             \else{\hbox{\kern.3em\inbar\kern-.3em{\rm C}}}\fi}
\def\IP{\rlx{\rm I\kern-.18em P}}
\def\IR{\rlx{\rm I\kern-.18em R}}
\def\Ione{\rlx{\rm 1\kern-2.7pt l}}
 %
%%%%%%%%%%%%%%%%%%%%%%%%%%%%%%%%%%%%%%%%%%%%%%%%%%%%%%%%%%%%%%%%%%%%%%%%
%%%****SHAPE****SHAPE****SHAPE****SHAPE****SHAPE****SHAPE****SHAPE****%%
 %
 % Get in shape, Man, (never mind the content)!

 %

\def\intem#1{\par\leavevmode%
              \llap{\hbox to\parindent{\hss{#1}\hfill~}}\ignorespaces}
 %
 % Indents #1 lines by width of #2 and puts #2 in the "niche".

 % A 2-column, \hangindent "niche":

 %
 % Math...
 % My version of \eqalign, \eqalignno ...
\newskip\humongous \humongous=0pt plus 1000pt minus 1000pt   % isn't it?
\def\caja{\mathsurround=0pt}
\newif\ifdtup
 %
 % display pattern: [         >a<         \cr]

 %
 % display pattern: [       >a &= b<      \cr]
\def\eqalign#1{\,\vcenter{\openup2\jot \caja
     \ialign{\strut \hfil$\displaystyle{##}$&$
      \displaystyle{{}##}$\hfil\crcr#1\crcr}}\,}
 %
 % display pattern: [ >a &= b< & >c &= d< \cr]

 %
 % display to full hsize, numbered at far right

 %
 % display pattern: [         >a<        &(*) \cr]

 %
 % display pattern: [       >a &= b<      &(*) \cr]

 %
 % display pattern: [    >a &= b &= c<    &(*) \cr]

 %
 % display pattern: [ >a &= b< & >c &= d< &(*) \cr]

 %
 % display pattern: [ >a &= b< &= >c &= d< &(*) \cr]

 %
 % For extra v-space between rows of a matrix or eqn-alignment,
 % use "\noalign{\vskip2mm}". In the above equation alignments,
 % "\openup2mm" does the same, but has no effect in "\matrix".
 %
%%%%%%%%%%%%%%%%%%%%%%%%%%%%%%%%%%%%%%%%%%%%%%%%%%%%%%%%%%%%%%%%%%%%%%%%
%%%****SHORT****SHORT****SHORT****SHORT****SHORT****SHORT****SHORT****%%
 %
 % Redefinitions of TeX's commands :
 %
          % Polish l-slash, L-slash
        % Scandinavian o-slash, O-slash
          % "paragraph" and "section" symbols
        % tie-after, cedilla
          % plain text dotless "i" and "j"
          % under-bar, under-dot in plain text
\def\,{\hskip1.5pt}           % a little space, in text and math mode
 %
 % Some abbreviations that save typing :
\let\a=\alpha
\let\b=\beta

\let\e=\epsilon     
                       \let\F=\Phi
\let\g=\gamma

                         \let\P=\Pi
\let\q=\theta                   \let\Q=\Theta
         
                   \let\S=\Sigma

                                       \let\X=\Xi

 %
 % Additional math symbols
 %
\def\Box{{\sqcap\mkern-12mu\sqcup}}
\def\lapp{\lower.4ex\hbox{\rlap{$\sim$}} \raise.4ex\hbox{$<$}}
\def\gapp{\lower.4ex\hbox{\rlap{$\sim$}} \raise.4ex\hbox{$>$}}
\def\con{\ifmmode\raise.1ex\hbox{\bf*}
          \else\raise.1ex\hbox{\bf*}\fi}

\def\dual{\relax\leavevmode\lower.9ex\hbox{\titlerms*}}
\def\define{\buildrel\rm def\over =}
\let\id=\equiv
\let\8=\otimes
 %
 %
%%%%%%%%%%%%%%%%%%%%%%%%%%%%%%%%%%%%%%%%%%%%%%%%%%%%%%%%%%%%%%%%%%%%%%%%
%%****MACROS***MACROS***MACROS***MACROS***MACROS***MACROS***MACROS****%%
 %
 % Math macros
 %
\let\ba=\overline
\let\2=\underline

\let\Tw=\widetilde
 %
 % Let's take arguments...
%  Use:  "A\like{B}".
\def\dt#1{{\buildrel{\smash{\lower1pt\hbox{.}}}\over{#1}}}
\def\pd#1#2{{\partial#1\over\partial#2}}

\def\6(#1){\relax\leavevmode\hbox{\eightrm(}#1\hbox{\eightrm)}}
\def\0#1{\relax\ifmmode\mathaccent"7017{#1}     % a little circle atop,
                \else\accent23#1\relax\fi}      % as a halo of a saint
\def\7#1#2{{\mathop{\null#2}\limits^{#1}}}      % puts #1 atop #2
\def\5#1#2{{\mathop{\null#2}\limits_{#1}}}      % puts #1 beneath #2
 %
 % Will grow vertically with size of argument
\def\bra#1{\left\langle #1\right|}
\def\ket#1{\left| #1\right\rangle}
\def\V#1{\langle#1\rangle}

 %
 % Vertical arrows with labels

 %
 % For vert. arrows to grow, say "\bigg\down\crlap{...}", using the
 % side-script: a label for vertical delimiters

 %

 %
 % Horizontal arrows that can grow
\newbox\t@b@x
\def\rightarrowfill{$\m@th \mathord- \mkern-6mu
     \cleaders\hbox{$\mkern-2mu \mathord- \mkern-2mu$}\hfill
      \mkern-6mu \mathord\rightarrow$}
\def\tooo#1{\setbox\t@b@x=\hbox{$\scriptstyle#1$}%
             \mathrel{\mathop{\hbox to\wd\t@b@x{\rightarrowfill}}%
              \limits^{#1}}\,}
\def\leftarrowfill{$\m@th \mathord\leftarrow \mkern-6mu
     \cleaders\hbox{$\mkern-2mu \mathord- \mkern-2mu$}\hfill
      \mkern-6mu \mathord-$}
\def\froo#1{\setbox\t@b@x=\hbox{$\scriptstyle#1$}%
             \mathrel{\mathop{\hbox to\wd\t@b@x{\leftarrowfill}}%
              \limits^{#1}}\,}
 %
 % fractions
\def\frac#1#2{{#1\over#2}}
\def\frc#1#2{\relax\ifmmode{\textstyle{#1\over#2}} % A small fraction,
                    \else$#1\over#2$\fi}           % good in text.
                            % Like {1\over{#1}}
 %
 % The basic Theorem-like macro, uses equation numbers
 % Use:  \Claim\cLABEL{Theorem}{This is a theorem.}
\def\Claim#1#2#3{\bigskip\begingroup%
                  \xdef #1{\secsym\the\meqno}%
                   \writedef{#1\leftbracket#1}%
                    \global\advance\meqno by1\wrlabeL#1%
                     \noindent{\bf#2}\,#1{}\,:~\sl#3\vskip1mm\endgroup}

\def\QED{\rlx\hfill$\Box$\kern-7pt\raise3pt\hbox{$\surd$}\bigskip}
 %
 % Math miscellanea
 %
\def\1{\raise1pt\hbox{,}}     % for the derivative comma amidst indices

\def\:{\buildrel!\over=}

\def\CP#1{\rlx\ifmmode\IP^{#1}\else\IP$^{#1}$\fi}
\def\cP#1{\rlx\ifmmode\IC{\rm P}^{#1}\else$\IC{\rm P}^{#1}$\fi}

\def\sll#1{\rlx\rlap{\,\raise1pt\hbox{/}}{#1}}
\def\Sll#1{\rlx\rlap{\,\kern.6pt\raise1pt\hbox{/}}{#1}\kern-.6pt}

\let\ttt=\textstyle

 %
 % Text miscellanea
 % By Knuth, use commas if interjected! As in: ..., \ie, ...
        % "exempli gratia", not "example given"
\def\ie{\hbox{\it i.e.}}        % "idem est"
       % "et cettera"
 %

\def\CY{Calabi-\kern-.2em Yau}

\def\3{\ifmmode\ldots\else$\ldots$\fi}
\def\\{\hfill\break}
\def\Z{\hfil\break\rlx\hbox{}\quad}
\def\3{\ifmmode\ldots\else$\ldots$\fi}

 %
 % References
 %

 %

 %

\def\NP#1{{\it Nucl.\,Phys.\,}{\bf#1\,}}
\def\PL#1{{\it Phys.\,Lett.\,}{\bf#1\,}}
\def\PRp#1{{\it Phys.\,Rep.\,}{\bf#1\,}}

\baselineskip=13.0861pt plus2pt minus1pt
\parskip=\medskipamount
\let\ft=\foot
\noblackbox
 %
 % For European paper standard
\def\Afour{\ifx\answ\bigans
            \hsize=16.5truecm\vsize=24.7truecm
             \else
              \hsize=24.7truecm\vsize=16.5truecm
               \fi}
 %
%%%%%%%%%%%%%%%%%%%%%%%%%%%%%%%%%%%%%%%%%%%%%%%%%%%%%%%%%%%%%%%%%%%%%%%%
%%============>>>>            SAVE  TIMBER            <<<<============%%
 %
\def\SaveTimber{\abovedisplayskip=1.5ex plus.3ex minus.5ex
                \belowdisplayskip=1.5ex plus.3ex minus.5ex
                \abovedisplayshortskip=.2ex plus.2ex minus.4ex
                \belowdisplayshortskip=1.5ex plus.2ex minus.4ex
                \baselineskip=12pt plus1pt minus.5pt
 \parskip=\smallskipamount
 \def\ft##1{\unskip\,\begingroup\footskip9pt plus1pt minus1pt\setbox%
             \strutbox=\hbox{\vrule height6pt depth4.5pt width0pt}%
              \global\advance\ftno by1
               \footnote{$^{\the\ftno)}$}{\ninepoint##1}%
                \endgroup}}
\catcode`@=12
%%%%%%%%%%%%%%%%%%%%%%%%%%%%%%%%%%%%%%%%%%%%%%%%%%%%%%%%%%%%%%%%%%%%%%%%
%
% Free: 
% \A \B \C \E \H \I \K \M \N \O \R \T \U \V \1 \9
%
%                          End of  "zyp.tex".                          %
%                        May the article begin!                        %
%
%%%%%%%%%%%%%%%%%%%%%%%%%%%%%%%%%%%%%%%%%%%%%%%%%%%%%%%%%%%%%%%%%%%%%%%%
%\draftmode
\sequentialequations
\SaveTimber  % packs the article 10% denser
 %
%%%%%%%%%%%%%%%%%%%%%%%%%%%%%%%%%%%%%%%%%%%%%%%%%%%%%%%%%%%%%%%%%%%%%%%%%%%%
%%%%%%%%%%%%%%%[ Some more or less useful definitions ]%%%%%%%%%%%%%%%%%%%%%
%%%%%%%%%%%%%%%%%%%%%%%%%%%%%%%%%%%%%%%%%%%%%%%%%%%%%%%%%%%%%%%%%%%%%%%%%%%%
 %
\def\rd{{\rm d}}
\def\hc{\hbox{\it h.c.}}

\def\@{{\ttt\cdot}}
                % or \def\pmp{{\pm\pm}}
                % or \def\mpm{{\mp\mp}}
 % or \def\pp{{++}}
                                % or \def\mm{{--}}

\def\]#1{\mkern#10mu}
\def\[#1{\mkern-#10mu}
\def\N{\nabla}
\def\bN{\overline\nabla}
 %

 %

 % (\lambda/2\pi)

\def\pB{\relax\leavevmode\hbox{$\BMfont p$\kern-.45em
               \vrule height1.35ex depth-1.25ex width4pt}\kern1pt}
\def\qB{\relax\leavevmode\hbox{$\BMfont q$\kern-.45em
               \vrule height1.35ex depth-1.25ex width4pt}\kern1pt}

\def\Cb{\relax\leavevmode\hbox{$C$\kern-.53em
               \vrule height1.9ex depth-1.8ex width5pt}\kern.5pt}
\def\Db{\relax\leavevmode\hbox{$D$\kern-.6em
               \vrule height1.9ex depth-1.8ex width5pt}\kern.5pt}
\def\rDb{\relax\leavevmode\hbox{D\kern-.7em
               \vrule height1.9ex depth-1.8ex width5pt}\kern1.5pt}
\def\Fb{\relax\leavevmode\hbox{$F$\kern-.55em
               \vrule height1.9ex depth-1.8ex width4.5pt}\kern1pt}
\def\FB{\relax\leavevmode\hbox{$\Phi$\kern-.6em
               \vrule height1.9ex depth-1.8ex width4.5pt}\kern1.5pt}
\def\Gb{\relax\leavevmode\hbox{$G$\kern-.55em
               \vrule height1.9ex depth-1.8ex width4.5pt}\kern1pt}

\def\JB{\relax\leavevmode\hbox{$\Psi$\kern-.7em
               \vrule height1.9ex depth-1.8ex width6pt}\kern1.5pt}
\def\LB{\relax\leavevmode\hbox{$\Lambda$\kern-.6em
               \vrule height1.9ex depth-1.8ex width5pt}\kern1.5pt}
\def\Mb{\relax\leavevmode\hbox{$M$\kern-.8em
               \vrule height1.9ex depth-1.8ex width7pt}\kern.8pt}
\def\Pb{\relax\leavevmode\hbox{$P$\kern-.55em
               \vrule height1.9ex depth-1.8ex width4.5pt}\kern.5pt}
\def\PB{\relax\leavevmode\hbox{$\P$\kern-.6em
               \vrule height1.9ex depth-1.8ex width4.5pt}\kern.5pt}
\def\Qb{\relax\leavevmode\hbox{$Q$\kern-.6em
               \vrule height1.9ex depth-1.8ex width5.5pt}\kern.5pt}
\def\QB{\relax\leavevmode\hbox{$\Q$\kern-.65em
               \vrule height1.9ex depth-1.8ex width5.0pt}\kern1.5pt}
\def\Tb{\relax\leavevmode\hbox{$T$\kern-.55em
               \vrule height1.9ex depth-1.8ex width4.5pt}\kern.5pt}
\def\Xb{\relax\leavevmode\hbox{$X$\kern-.67em
               \vrule height1.9ex depth-1.8ex width6pt}\kern.5pt}
\def\XB{\relax\leavevmode\hbox{$\X$\kern-.65em
               \vrule height1.95ex depth-1.85ex width6pt}\kern.5pt}
\def\YB{\relax\leavevmode\hbox{$\Upsilon$\kern-.6em
               \vrule height1.9ex depth-1.8ex width4.5pt}\kern1.5pt}
\def\Wb{\relax\leavevmode\hbox{$W$\kern-.9em
               \vrule height1.9ex depth-1.8ex width7pt}\kern1.5pt}

\def\cDb{\relax\leavevmode\hbox{$\cal D$\kern-.6em
               \vrule height1.9ex depth-1.8ex width5pt}\kern.5pt}
\def\ad{{\dot\a}}
\def\bd{{\dot\b}}

\def\bAb{\relax\leavevmode\hbox{{\bf A}\kern-.7em
               \vrule height1.9ex depth-1.8ex width5pt}\kern2pt}

\def\bBb{\relax\leavevmode\hbox{{\bf B}\kern-.7em
               \vrule height1.9ex depth-1.8ex width5pt}\kern2pt}

\def\rQb{\relax\leavevmode\hbox{Q\kern-.65em
               \vrule height1.9ex depth-1.8ex width5pt}\kern1.5pt}
\def\bQb{\relax\leavevmode\hbox{{\bf Q}\kern-.65em
               \vrule height1.9ex depth-1.8ex width5pt}\kern1.5pt}

\def\ssl#1{\rlx\rlap{\,\raise1pt\hbox{$\backslash$}}{#1}}
\def\Ssl{\rlap{\kern1.2pt\raise1pt\hbox{\rm/}}{\hbox{$S$}}}

 %

 %
 % For displaying captioned EPSF figures with epsf.tex

 %
%%%%%%%%%%%%%%%%%%%%%%%%%%%%%%%%%%%%%%%%%%%%%%%%%%%%%%%%%%%%%%%%%%%%%%%%%%%%
%%%%%%%%%%%%%%%%%%%%%%%%%%%[ Here we go ! ]%%%%%%%%%%%%%%%%%%%%%%%%%%%%%%%%%
%%%%%%%%%%%%%%%%%%%%%%%%%%%%%%%%%%%%%%%%%%%%%%%%%%%%%%%%%%%%%%%%%%%%%%%%%%%%
 %
\Title{\rightline{hep-th/9903175}}
      {\vbox{\centerline{Linear and Chiral Superfields}
              \vskip2mm
             \centerline{are Usefully Inequivalent}}}
\centerline{\titlerms Tristan H\"ubsch\footnote{$^{\spadesuit}$}{On leave
            from the ``Rudjer Bo\v skovi\'c'' Institute, Zagreb, Croatia.}}
                                                             \vskip0mm
 \centerline{\it Department of Physics and Astronomy}        \vskip-.5mm
 \centerline{\it Howard University, Washington, DC~20059}    \vskip-.5mm
 \centerline{\tt thubsch\,@\,howard.edu}
\vfill

\centerline{ABSTRACT}\vskip2mm
\vbox{\narrower\narrower\baselineskip=12pt\noindent
Chiral superfields have been used, and extensively, almost ever since
supersymmetry has been discovered. Complex linear superfields afford an
alternate representation of matter, but are widely misbelieved to be
`physically equivalent' to chiral ones. We prove the opposite is true.
Curiously, this re-enables a previously thwarted interpretation of the
low-energy (super)field limit of superstrings.}
 \vskip+5mm
 \rightline{{\ninepoint\it Brevity is the soul of wit.}}
 \rightline{{\ninepoint\it|~W.~Shakespeare}}

\Date{March '99. \hfill}  % the real thing
\footline{\hss\tenrm--\,\folio\,--\hss}
 %
 %
%%%%%%%%%%%%%%%%%%%%%%%%%%%%%%%%%%%%%%%%%%%%%%%%%%%%%%%%%%%%%%%%%%%%%%%%%%%%
%%%%%%%%%%%%%%%%%%%%%%%%%%%[ References ! ]%%%%%%%%%%%%%%%%%%%%%%%%%%%%%%%%%
%%%%%%%%%%%%%%%%%%%%%%%%%%%%%%%%%%%%%%%%%%%%%%%%%%%%%%%%%%%%%%%%%%%%%%%%%%%%
 %
%%%% Books, reviews & lectures
 %
\lref\rAKMRV{D.~Amati, K.~Konishi, Y.~Meurice, G.C.~Rossi and
       G.~Veneziano:\Z \PRp{162C}(1988)170, and references therein.}

\lref\rBK{I.L.~Buchbinder and S.M.~Kuzenko: {\it Ideas and Methods of
        Supersymmetry and\Z Supergravity : Or a Walk Through Superspace},
        (IOP, Bristol, 1998).}

\lref\rGGRS{S.J.~Gates, Jr., M.T.~Grisaru, M.~Ro\v cek and
       W.~Siegel: {\it Superspace}\Z (Benjamin/Cummings Pub.\ Co.,
       Reading, Massachusetts, 1983).}

\lref\rGSW{M.B.~Green, J.H.~Schwarz and E.~Witten: {\it Superstring
       Theory II}\Z (Cambridge University Press, Cambridge, 1987).}

\lref\rBeast{T.~H\"ubsch: {\it \CY\ Manifolds---A Bestiary for
       Physicists}\Z (World Scientific, Singapore, 1992).}

\lref\rMKW{H.~Muller-Kirsten and A.~Wiedmann: {\it Supersymmetry:
        An Introduction With\Z Conceptual and Calculational Details}
        (World Scientific, Singapore, 1987).}

\lref\rWB{J.~Wess and J.~Bagger: {\it Supersymmetry and Supergravity}\Z
      (Princeton University Pub., Princeton NJ, 1983).}

\lref\rPW{P.~West: {\it Introduction to Supersymmetry and Supergravity}\Z
      (World Scientific, Singapore, 2nd ext,~ed.: 1990).}

%%%% Articles
 %
\lref\rNMChJD{S.J.~Gates, Jr.\ and B.B.~Deo: \NP{B254}(1985)187-200.}

\lref\rNMChJW{S.J.~Gates, Jr.\ and W.~Siegel: \NP{B187}(1981)389.}

\lref\rTwJim{S.J.~Gates, Jr.: \PL{B352}(1995)43--49.}

\lref\rUseNM{S.J. Gates, Jr.\ and Sergei M. Kuzenko: The
       CNM-Hypermultiplet Nexus.\Z hep-th/9810137.}

\lref\rGHR{S.J.~Gates, Jr., C.M.~Hull and M.~Ro\v cek: \NP{B248}(1984)157.}

\lref\rSSQP{T.~H\"ubsch, H.~Nishino and J.C.~Pati: \PL{163B}(1985)111. }

\lref\rHSS{T.~H\"ubsch: Haploid (2,2)-Superfields In 2-Dimensional
      Spacetime. hep-th/9901038.}

 %
%%%%%%%%%%%%%%%%%%%%%%%%%%%%%%%%%%%%%%%%%%%%%%%%%%%%%%%%%%%%%%%%%%%%%%%%%%%%
%%%%%%%%%%%%%%%%%%%%%%%%%%%[ The Article! ]%%%%%%%%%%%%%%%%%%%%%%%%%%%%%%%%%
%%%%%%%%%%%%%%%%%%%%%%%%%%%%%%%%%%%%%%%%%%%%%%%%%%%%%%%%%%%%%%%%%%%%%%%%%%%%
 %
\secno=-1
\newsec{Introduction}\noindent
Chiral superfields and their conjugates have been used for over two
decades~\refs{\rWB,\rGGRS,\rPW,\rMKW,\rBK}.
They are defined by a system of two homogeneous,
simple, first order superdifferential equations:
\eqn\eCh{ [\Db_\ad,\F]=0~,\qquad [D_\a,\FB]=0~, }
written here in $N{=}1$ supersymmetric 4-dimensional spacetime notation.
The complex linear superfield~\refs{\rNMChJW,\rNMChJD}, $\Q$, and its
conjugate, $\QB$, are defined to obey the homogeneous, single, second order
superdifferential equations:
\eqn\eLn{ [\Db^2,\Q]=0~,\qquad [D^2,\QB]=0~, }
where~\refs{\rWB}:
\eqn\eXXX{ D^2=D^\a D_\a=\e^{\a\b}D_\b D_\a~, \quad\hbox{and}\quad
           \Db^2=D_\ad\Db^\ad=\e^{\ad\bd}\Db_\ad\Db_\bd~.}
The complex linear superfield is also called `non-minimal chiral
superfield', as its defining equation is closely related to those of the
chiral superfield~\eCh; in retrospect, $\F$ ought to be called `minimal
chiral'\refs{\rHSS}. For the purposes of this short note, we adopt the
shorter, `chiral' and `linear' names, and moreover use only complex such
fields.

The simple Lagrangian densities
\eqn\eFree{ L^{(0)}_\F=\int\rd^4\q~\F\FB~,\quad\hbox{and}\quad
            L^{(0)}_\Q=-\int\rd^4\q~\Q\QB~, }
rewritten in terms of the component fields are in a 1--1 correspondence
after the elimination of auxiliary component fields~\refs{\rGGRS}. Even
in the current literature, this fact has been misinterpreted into a `proof'
that chiral and linear superfields are physically equivalent.

Herein, we dispell such and related delusions by offering two simple
proofs, which rely {\it only on the definition\/} of these superfields,
and focus on two very distinct aspects of the inequivalence between them.
These proofs encompass, {\it mutatis mutandis\/}, all of the `minimal' and
`non-minimal' haploid superfields of Ref.~\refs{\rHSS}.

\newsec{1st Proof: Correlation Functions}\noindent
\seclab\sFrst
It is a rather well exploited fact that chiral superfields form a
ring. In fact, any analytic function of a collection of chiral
superfields is also a chiral superfield.

This, rather obviously, is not the case with linear superfields: no
nonlinear function of even a single linear superfield is linear, \ie, obeys
Eqs.~\eLn. That is, the set of linear superfields is not closed under
multiplication: the product $\Q_1\Q_2$ does not obey Eq.~\eLn.

This simple mathematical fact has an extraordinary consequence when
comparing {\it any two models\/} using chiral {\it vs.\/}\ linear
superfields. Supersymmetric correlation functions of chiral superfields are
always independent of the (spacetime) position of
the superfields~\refs{\rAKMRV}:
\eqn\eRgd{{\eqalign{\pd{}{x^m_1}\V{*|\F_1\F_2|*}
 &=\Big\langle*\Big|\Big(\pd{\F_1}{x^m_1}\Big)\F_2\Big|*\Big\rangle
  \propto\g_m^{\a\ad}\V{*|\big\{\Qb_\ad,[Q_\a,\F_1]\big\}\F_2|*}~, \cr
 &\propto\g_m^{\a\ad}\underbrace{\bra{*}\Qb_\ad}_{\id0}[Q_\a,\F_1]\F_2
   +[Q_\a,\F_1]\F_2\underbrace{\Qb_\ad\ket{*}}_{\id0}~\id~0~, \cr
 }}}
where $\ket{*}$ is any supersymmetric vacuum, annihilated by all the
supercharges, and we used the defining equations~\eCh. Thus, models
involving chiral superfields always have a `topological' (more properly,
rigid) sector among their correlation functions.

By stark contrast and {\it as a consequence of their differing definition
only\/}, no such rigidity obtains for supersymmetric correlation functions
of products of linear superfields: models built with linear superfields in
place of chiral ones have no such `topological' sector.

As correlation functions determine a (quantum) model, it follows that
models with linear superfields are fundamentally inequivalent to
models with chiral superfields.

\newsec{2nd Proof: Gauge Interactions, in 2 Dimensions}\noindent
\seclab\sScnd
Recall that Eqs.~\eCh\ and~\eLn\ hold also in 2-dimensions, but the Lorentz
group now becoming $SL(2,\IC)\to SO(1,1)\approx U(1)$, we drop the over-dots
on the spinorial indices.

Consider coupling these superfields to a gauge symmetry. As shown in
Ref.~\refs{\rTwJim}, 2-dimensional supersymmetric theories admit a twisted
version of the usual gauge vector multiplet the covariant (super)derivatives
of which satisfy $(\hat{m}=1,2)$:
\eqn\eVCh{ \{\N_\a,\N_\b\}=4ig\g^3_{\a\b}\Pb~,\qquad
 \{\N_\a,\bN_\b\}=2i\g^{\hat{m}}_{\a\b}\N_{\hat{m}}~,
}
where $\Pb$ is the Lie algebra valued superfield the superderivatives of
which provide the twisted versions of the usual fermionic and bosonic field
strengths and the auxiliary field:
\eqn\eXXX{ [\bN_\a,\Pb] = \Tw{\ba{W}}_\a~,\qquad
           [\N_\a,\Tw{W}_\b] = \e_{\a\b}\big(\Tw{F}+i\Tw{{\rm D}}\big)~. }
In the presence of a gauge symmetry, the definitions~\eCh\ and~\eLn\ need
to be modified into
\eqn\eChG{ [\bN_\a,\F]=0~,\qquad [\N_\a,\FB]=0~, }
and
\eqn\eLnG{ [\bN^2,\Q]=0~,\qquad [\N^2,\QB]=0~. }
Note that, as defined above, $\Pb$ automatically satisfies the latter
of Eqs.~\eChG; its conjugate, $P$ is then chiral, whence we call
$\{P,\Tw{W}_\a,\Tw{F},\Tw{{\rm D}}\}$ the `chiral vector multiplet'.
The dimensionally reduced standard gauge multiplet is likewise determined
from a twisted chiral superfield~\refs{\rTwJim}, $\S$, and we call
$\{\S,W_\a,F,{\rm D}\}$ the `twisted gauge vector multiplet'.

There is an `integrability' condition one can derive for the chiral
superfields~\refs{\rTwJim}, by anticommuting $[\bN_\a,\F]$ in~\eChG with
$\bN_\b$, and antisymmetrizing:
\eqn\eXXX{
 0=\big\{\bN_\a,[\bN_\b,\F]\big\}+\big\{\bN_\b,[\bN_\a,\F]\big\}=
 \big[\{\bN_\a,\bN_\b\},\F\big]\define4ig\g^3_{\a\b}[\Pb,\F]~. }
This implies that {\it covariantly\/} chiral superfields~\eChG\ must be
{\it chargeless\/} with respect to the gauge group generated by the chiral
gauge vector multiplet $\{P,\Tw{W}_\a,\Tw{F},\Tw{{\rm D}}\}$.

Again, {\it as a consequence of their differing definition only\/}, no
such restriction obtains for the {\it covariantly\/} linear
superfield~\eLnG.

The `mirror' of this argument proves that superfields which are
{\it covariantly\/} twisted-chiral with respect to the twisted gauge
multiplet must also be chargeless with respect to this gauge symmetry.

To sum up, covariantly chiral superfields couple to the twisted but not to
the chiral gauge multiplet~\eVCh, while covariantly twisted-chiral
superfields couple to the chiral but not to the twisted vector multiplet.
By stark contrast, both linear superfields and their `mirror'-twisted
brethren couple indiscriminately to both chiral and twisted gauge
multiplets, and so cannot possibly be equivalent to the discriminately
coupling (twisted-)chiral ilk.

\newsec{An Old Proof Opens New Possibilities}\noindent
\seclab\sThrd
An unassuming remark in Ref.~\refs{\rGGRS}, p.\,200, states: ``it is
not possible to introduce arbitrary mass and nonderivative self-interaction
terms'' for the linear complex superfield.

This is clear from the fact that the definition of a chiral superfield~\eCh\
ensures that {\it its most general\/} Lagrangian density takes the form:
\eqn\eChL{ L_\F=\int\rd^4\q~K(\F,\FB) + \int\rd^2\q~W(\F) ~+~ \hc~, }
whereas the definition of a linear superfield~\eLn\ restricts {\it its
most general\/} Lagrangian to:
\eqn\eLnL{ L_\Q=\int\rd^4\q~f(\Q,\QB) ~+~ \hc }
Besides the sophisticated consequences for
renormalization\ft{$\int\rd^4\q$-terms notoriously undergo renormalization,
and generally in a complicated fashion, whereas $\int\rd^2\q$-terms {\it
typically\/} do not: see p.~358 of Ref.~\refs{\rGGRS}.}, an important
difference between~\eChL\ and~\eLnL\ is based {\it on dimensional grounds
only\/}: when restricted to be renormalizable, in 4 dimensions,
$L_\Q$ admits {\it no mass\/} parameter and {\it must be quadratic\/}!

This proves that renormalizable models with (only) linear superfields
cannot describe massive fields with Yukawa interactions, whereas
renormalizable models with chiral superfields can. The two then cannot
possibly be equivalent|even in the free field limit!

This brings about an interesting possibility. Recall that the effective
low-energy (super)field limit of superstring theories~\refs{\rGSW,\rBeast}
is identified based only on physical (propagating) degrees of freedom. It
is then {\it logically possible} that the limit in fact ought to involve
{\it linear\/} instead of {\it chiral\/} superfields. If so, matter fields
descending from superstrings would be massless and have only gauge
interactions. This is precisely as needed for the models discussed in
Ref.~\refs{\rSSQP}, which when using chiral superfields end up thwarted
mainly by Yukawa intereactions.

Note also that the {\it combined\/} use of chiral and linear superfields
enables the description of phenomena not describable by the use
of either one of the superfields alone~\refs{\rUseNM}. Indeed, owing to no
self-interaction, linear superfields are well adapted to describe the
flatness of (co)tangent spaces. This is not unlike the richness in
2-dimensional (super)spacetime, made possible by the {\it joint\/} use of
chiral and twisted-chiral superfields in Ref.~\refs{\rGHR}, and a far
greater richness made possible by the use of many other types of
superfields~\refs{\rHSS}.

%%%%%%%%%%%%%%%%%%%%%%%%%%%%%%%%%%%%%%%%%%%%%%%%%%%%%%%%%%%%%%%%%%%%%%%%%%%%
%%%%%%%%%%%%%%%%%%%%%%[ Long live our  Benefactors ! ]%%%%%%%%%%%%%%%%%%%%%%
%%%%%%%%%%%%%%%%%%%%%%%%%%%%%%%%%%%%%%%%%%%%%%%%%%%%%%%%%%%%%%%%%%%%%%%%%%%%
 %
\vfill%\eject
{\ninepoint\bigskip\noindent{\it Acknowledgments\/}:
Many thanks to S.J.~Gates, Jr.\ for helpful discussions, to the
generous support of the Department of Energy through the grant
DE-FG02-94ER-40854,
  and also to the Institute for Theoretical Physics at Santa
  Barbara, where part of this work was done with the support from the
  National Science Foundation, under the Grant No.~PHY94-07194.}
\vfill%\eject

\bigskip\listrefs

%%%%%%%%%%%%%%%%%%%%%%%%%%%%%%%%%%%%%%%%%%%%%%%%%%%%%%%%%%%%%%%%%%%%%%%%%%%%
%%%%%%%%%%%%%%%%%%%%%%%%%%%%[ Goody-Good-Bye ! ]%%%%%%%%%%%%%%%%%%%%%%%%%%%%
%%%%%%%%%%%%%%%%%%%%%%%%%%%%%%%%%%%%%%%%%%%%%%%%%%%%%%%%%%%%%%%%%%%%%%%%%%%%
 %
\bye